\documentclass[%
 reprint,
 amsmath,amssymb,
 aps,
]{revtex4-2}

\usepackage{graphicx}% Include figure files
\usepackage{dcolumn}% Align table columns on decimal point
\usepackage{bm}% bold math
\usepackage{xcolor}
\newcommand{\QFI}{\mathcal{J}}
\newcommand{\mean}[1]{\left\langle#1\right\rangle}

\newcommand{\nbraket}[2]{\langle #1| #2\rangle}
\newcommand{\xiQL}{\xi_{\text{QL}}}
\newcommand{\xiNOON}{\xi_{\text{NOON}}}
\newcommand{\xiSP}{\xi_{\text{SP}}}
\newcommand{\xiMP}{\xi_{\text{MP}}}
\newcommand{\xiSQZ}{\xi_{\text{SQZ}}}

\newcommand{\nsq}{n_{\text{sq}}}
\newcommand{\etart}{\eta_{\text{rt}}}
\newcommand{\etap}{\eta_{\text{P}}}
\newcommand{\etam}{\eta_{\text{D}}}

\begin{document}

\preprint{APS/123-QED}

\title{Optimal Dose-Limited Phase Estimation without Entanglement}% 

\author{Stewart A. Koppell}
 \email{skoppel2@jh.edu}
\author{Mark A. Kasevich}%

\affiliation{%
 Physics Department, Stanford University
}%

\date{\today}% It is always \today, today,

\begin{abstract}
Phase estimation is one of the central themes of quantum metrology, with applications in sensing, microscopy, and quantum computation. When estimating a phase shift in a lossy medium, there is an upper bound on the attainable information per probe particle particle sent through the phase shifting element. Previously, only entanglement-enhanced measurements have been shown to saturate this bound. We describe a measurement scheme which can saturate the bound without relying on entanglement.  
%<600 characters
\end{abstract}

\maketitle

\section{Introduction}
A central theme of quantum metrology is that entanglement in the probe state can enhance measurement precision. The NOON state interferometer, for example, achieves Heisenberg-limited precision using a maximally entangled $n$-particle probe \cite{Holland:1993}. While Heisenberg-limited measurements are often thought of as a uniquely quantum phenomenon, the same precision scaling is possible with a multi-pass (MP) measurement where a classical probe passed $m$ times through the interferometer. In some treatments, $n$ and $m$ are defined as equivalent resources and MP measurements are said to be Heisenberg-limited \cite{EntanglementfreeHL1,EntanglementfreeHL2,EntanglementfreeHL3}.  To clarify when a quantum advantage exists, it is important to define the physical measurement resources and limitations. 

When measuring a dynamical quantity such as in magnetometry or gravitational field detection, measurement bandwidth limits the maximum time the probe can coherently accrue its phase shift. Once this limit is saturated, the only way to increase sensitivity is with entanglement in the probe. In applications such as these, where $m$ has an upper limit, the Heisenberg limit is exclusive to quantum measurements.

Phase estimation is also used in quantum computing as a subroutine for many important algorithms (e.g. Shor's algorithm). Kitaev's phase estimation algorithm \cite{Kitaev:1996} and the iterative phase estimation algorithm (IPEA) \cite{Griffiths:1996,Dobsicek:2007} can achieve a precision scaling like $1/2^{nm}$ using $n$ ancillary qubits and $m\log{nm}$ iterations \cite{Dobsicek:2007}. Thus spatial resources (qubits) can be traded for temporal resources (runtime) while maintaining an exponential advantage. 

The discrimination of unitary operations is another noteworthy measurement task where entanglement and multi-passing can be compared. In general, any single-pass (SP) measurement with a separable probe state repeated a finite number of times has a non-zero probability of failure. Perfect discrimination was first described as an application of entanglement \cite{entangle_descrim_1,entangle_descrim_2}. Later, in a letter which posed the question ``What kind of tasks can be achieved without entanglement?", it was shown that a MP measurement can also suffice \cite{MPdescrim}.

In the presence of loss, MP and NOON-state measurements are no longer optimal for phase estimation. More general SP Fock-state measurements can outperform MP measurements \cite{DD2010}, and coherent states in a ring resonator outperform all SP quantum probes \cite{Jonathan2022}.

The primary interests of this article are applications like microscopy where the measurement is performed on a material sample which may be destroyed or altered by inelastic interactions with the probe. We assume the sample can be modeled as a lossy static phase shift and seek to minimize the measurement error per dose $d$. If $|\alpha_k|^2$ is the intensity incident on the sample at measurement stage $k$, then
\begin{equation}
    d=\sum_{k=0}^{m-1}\left|\alpha_k\right|^2\ .
\end{equation}
For SP interferometer measurements, the dose metric simply counts the number of particles which enter the sample channel. For MP measurements, the dose metric takes into account the losses on each pass, which reduce the dose inflicted on subsequent passes. In optical microscopy, both entanglement-enhanced measurements \cite{multiphaseEstPRL13,NOONmicroscopyPRL14,QuantumenhancedNM,Taylor2013,Crespi2012} and MP measurements \cite{Juffmann:2016,Nimmrichter2019} have been proposed to improve dose-limited imaging sensitivity. Similar schemes for transmission electron microscopes have been proposed \cite{Okamoto2013,Putnam2009,Juffmann:2017}. However it has remained unclear whether entanglement is necessary to achieve the best possible measurement sensitivity per dose.

To answer this question, we will employ the Quantum Fisher Information (QFI) formalism. Via the Quantum Cramer Rao Bound (QCRB), the QFI provides a lower limit on the error $\Delta_\theta$ in an estimate of an unknown parameter $\theta$ \cite{Holevo1982,Helstrom1967,Braunstein1994}. The QCRB for a measurement repeated $N$ is
\begin{equation}
    \Delta_\theta^2\geq\frac{1}{N\QFI}\ .
\end{equation}
where $\mathcal{J}$ is the QFI. The QFI for a lossless NOON-state interferometer is proportional to $n^2$. This quadratic scaling, characteristic of the Heisenberg limit, is not the only hallmark of quantum advantage. In fact, in the presence of any amount of loss, the QFI cannot asymptotically scale superlinearly in $n$ \cite{QL,DD:2012}. Instead, quantum measurements may confer a constant advantage over the classical alternatives. 

With dose as the measurement resource, the figure of merit to optimize is $\xi=\QFI/d$, which we will refer to as the dose efficiency. For a sample with transmissivity $\eta$, the quantum limit for dose efficiency is \cite{QL,Birchall2017}
\begin{equation}
    \xiQL=\frac{4\eta}{(1-\eta)}\ .
\end{equation}
It is important to note that the optimal measurement typically depends on the unknown value of $\theta$. For example, Zernike phase contrast \cite{Zernike1942} is optimal for classical SP imaging of weakly-scattering objects, but ineffective for strong scatterers \cite{Koppell2021}. Saturating the QCRB is generally only possible asymptotically (for large $N$) with an adaptive measurement sequence or with significant \textit{a-priori} knowledge \cite{Hradil2005,Holevo1982}. In the context of microscopy, measurements typically involve many photons ($N\gg1$) and a strong \textit{a-priori} condition often applies (e.g. weak scatterers have a phase shift $|\theta(\vec{r})-\theta_0|\ll1$ for some constant $\theta_0$). However, our results will not rule out the possibility that quantum probes have superior performance for strong phase objects and small $N$.

%%%%%%%%%%%%%%%%%%%%%%%%%%%%%%%%%%%%%%%%%%%%
\section{Dose as a Measurement Resource}
\label{sec:sp}
%%%%%%%%%%%%%%%%%%%%%%%%%%%%%%%%%%%%%%%%%%%%

It will be instructive to begin by calculating $\xi$ for a SP Mach-Zehnder interferometer using classical, NOON-state, and squeezed-state probes. A standard approach to calculating the QFI in the presence of loss involves Kraus operators, which describe the dynamics of open quantum systems \cite{Dorner2009,DD:2009,DD:2015}. We can take a simpler approach when the probe is a pure single-particle state or a NOON state, as in these cases no information is available if there is any absorption. Then $\QFI=(1-p_{\text{abs}})\tilde{\QFI}$ where $\tilde{\QFI}$ is the QFI in the absence of absorption and $p_{\text{abs}}$ is the absorption probability. 

We begin with a lossy Mach-Zehnder interferometer (MZI) using a separable input state. If $\alpha$ is the amplitude sent to the sample arm, the normalized exit wavefunction is
\begin{equation}
    \psi(\theta)=\frac{1}{\sqrt{p}}
    \begin{bmatrix}
    \sqrt{1-\alpha^2}\\\alpha\sqrt{\eta}e^{i\theta}
    \end{bmatrix}\ .
\end{equation}
where $p=1-p_{\text{abs}}$ and $p_{\text{abs}}=\alpha^2(1-\eta)$ is the absorption probability. The QFI for a pure state is \cite{Holevo1982}
\begin{equation}
\QFI=4\nbraket{\dot{\psi}}{\dot{\psi}}-4|\nbraket{\dot{\psi}}{\psi}|^2\ ,
\end{equation}
where the dot indicates a derivative with respect to $\theta$. So the the lossy MZI has
\begin{equation}
\tilde{\QFI}=\frac{4}{p}|\alpha|^2\eta\left(1-\frac{1}{p}|\alpha|^2\eta\right)\ .
\end{equation}
The expected QFI per dose $d=|\alpha|^2$ is maximized in the strong local oscillator limit ($\alpha\rightarrow0$), yielding
\begin{equation}
    \xiSP=\frac{p\tilde{\QFI}}{d}=4\eta\ .
\end{equation} 
The principle that maximum dose efficiency requires a strong local oscillator will be true for each of the measurements we discuss below. Compared to the quantum limit, the MZI yields less information per dose by a factor of $1/(1-\eta)$, which diverges as $\eta\rightarrow 1$.

In the presence of loss, NOON state probes are sub-optimal because they fail when a single particle is absorbed. Nevertheless, an unbalanced NOON state in a MZI (with probability $|\alpha|^2\ll1$ of finding $n$ particles in the sample channel and probability $1-|\alpha|^2$ of finding $n$ particles in the reference channel) is a significant improvement over a separable probe. For $\eta\sim 1$, $\xiNOON/\xiQL\gtrsim 1/e$ for optimal $n$. Explicitly, $\QFI=4n^2\eta^n|\alpha|^2$ and $d=n|\alpha|^2$, so
\begin{equation}
    \xiNOON=4n\eta^n\ ,
\end{equation}
which is maximized by choosing $n\sim-1/\ln{\eta}$.

MP and NOON state interferometry provide the same QFI when $n=m$ and are both measurements spoiled by a single absorption. However, the dose from a MP measurement is limited: there can be no more than a one absorption. As a result, MP measurements are more dose efficient. For optimal $m$, $\xiMP/\xiQL\gtrsim 0.65$. In general $J=4m^2\eta^m|\alpha|^2$ and $d=\sum_{k=0}^{m-1}\eta^k|\alpha|^2$, so
\begin{equation}
    \xiMP=4m^2\eta^m\frac{1-\eta}{(1-\eta^m)}\ .
\end{equation}
While a MP interferometer does not fully achieve the quantum limit, the simplicity of the scheme may make it a practical choice in many applications.

Finally, the expected number of particles in a pure single mode squeezed Gaussian state is $\mean{n}=\alpha^2+\sinh^2(r)$ where $\alpha$ is the displacement parameter and $r$ is the squeezing parameter (measured in dB, the squeezing is $20\ln{r}/10$). The number of particles contributing to the squeezing is $\nsq=\sinh^2(r)$. If $\mean{n}\gg\nsq$, then \cite{Birchall2017}
\begin{equation}
    \xiSQZ(\nsq)=\frac{4\eta}{2\eta\nsq-2\eta\sqrt{(\nsq(\nsq+1))}+1}\ .
\end{equation}
When $\nsq$ is large, the denominator approaches $1-\eta$ and this measurement saturates the quantum limit for dose efficiency. 

%%%%%%%%%%%%%%%%%%%%%%%%%%%%%%%%%%%%%%%%%%%%
\section{Phase Estimation with a Chain Interferometer}
%%%%%%%%%%%%%%%%%%%%%%%%%%%%%%%%%%%%%%%%%%%%
\begin{figure*}
\includegraphics[width=\textwidth]{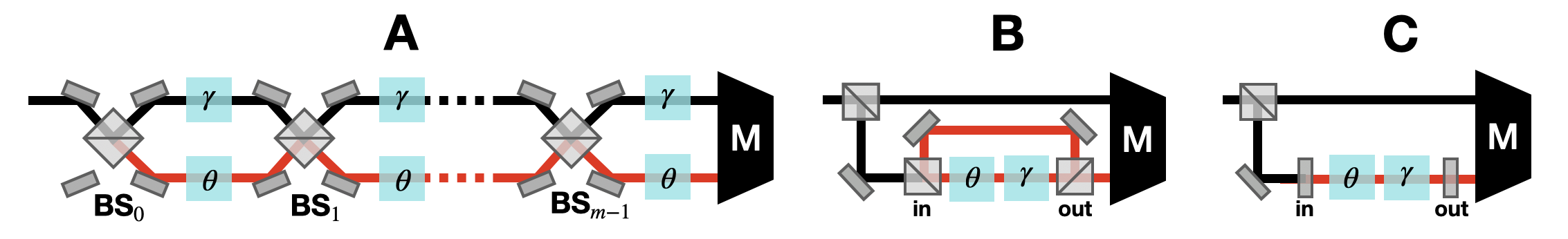}
\caption{\label{fig:schematic} (A) Chain interferometer (CI) scheme for estimation of an unknown phase $\theta$. At each stage $k$ of $m$, the reference channel (black) and sample channel (red) are weakly coupled by a beamsplitter (BS$_k$) with transmissivity $T_k\ll 1$ before the probe interacts with the unknown phase shift $\theta$ in the sample arm and a programmed phase shift $\gamma$ in the reference arm. A measurement $M$ is then performed on the output state. (B) If $T_k$ increases geometrically with $k$, the CI is equivalent to a ring cavity. (C) When $T_k$ increases geometrically for odd $k$ and $T_k\sim 0$ for even $k$, the CI is equivalent to a Fabry-Perot cavity. Both cavity interferometers are optimized by setting the in-coupling transmissivity to $T_\text{in}\ll1$, the out-coupling transmissivity equal the round-trip loss, and $\gamma$ to cancel the round-trip phase shift. The block labeled $M$ represents a homodyne measurement.}
\end{figure*}

Various applications have been found for interferometer schemes more sophisticated than the SP or MP MZI. A simple example is the double MZI, formed by placing two MZIs in series \cite{Ataman2014,Szigeti2020,Ham2021}. With a long chain of MZIs, it's possible to perform an interaction-free measurement (IFM), where the presence of an absorber in the sample channel can be detected with an arbitrarily small chance of absorption \cite{kwiat1995}. The IFM interferometer can be optimized for phase contrast by adjusting $m$, the beamsplitter transmissivity $T$, and the reference phase shift $\gamma$. These parameters have been numerically optimized in Ref. \cite{Birchall2017}, which found that with $T\approx (0.01/2m)^2$ and $\gamma=0$ (assuming small $\theta$), $\xi/\xiQL>0.9$. 

The most general possible $m$-stage entanglement-free phase estimation scheme could use up to $m$ auxiliary channels in addition to the sample channel, with unitary transformations applied to couple all the channels between each stage. This would likely be infeasible to implement as a microscope architecture. Remarkably, we will show that it is possible to saturate the QFI using a chain interferometer (CI) - a generalization of the IFM architecture where the beamsplitter transmissivities $T_k$ are allowed to change at each measurement stage. Fig. \ref{fig:schematic} (A) shows a CI with $m$ stages. 

To maintain a strong local oscillator, we will allocate the majority of the probe amplitude to the reference arm with the ansatz $T_k<\epsilon^2$ for some $\epsilon\ll 1$. We will also set $\gamma\sim\theta$ (this choice requires \textit{a-priori} knowledge of $\theta$, such as the weak scattering condition mentioned in the introduction). Unlike a MP interferometer, the exit wavefunction of a CI contains terms corresponding to different numbers of passes through the sample. Interference between these terms provides extra information. 

The details of calculating $\QFI$ and $d$ for the CI are given in the supplementary material (Sec. A). To order $\epsilon^2$, 
\begin{equation}
    \QFI=4\left|\sum_{k=0}^{m-1}(m-k)\eta^{(m-k)/2}\tau_k\right|^2
\end{equation}
and
\begin{equation}
    d=\sum_{k=0}^{m-1}\left|\sum_{k'=0}^{k}\eta^{(k-k')/2}\tau_{k'}\right|^2\ ,
\end{equation}
where $\tau_k=\sqrt{T_k}$. By choosing the values
\begin{equation}
    \tau_k=\begin{cases}
    \epsilon\eta^{m/2} & k=0\\
    \epsilon(1-\eta)\eta^{(m-k)/2} & k>0
    \end{cases}\ .
\end{equation}
we find
\begin{equation}
\lim_{m\rightarrow\infty}\QFI=\epsilon^2\frac{4\eta^2}{(1-\eta)^2}\quad \text{and}\quad \lim_{m\rightarrow\infty}d=\epsilon^2\frac{\eta}{1-\eta}\ .
\end{equation}
so that $\xi\rightarrow\xi_{QL}$ for $m\rightarrow\infty$. In that limit, the fraction of probe intensity passing through the sample exactly $k$ times is proportional $\eta^k$. 

For probes with long coherence times, an equivalent output state is formed by a leaky optical ring resonator as shown in Fig. \ref{fig:schematic} (B) with in-coupling transmissivity $\epsilon^2$ and an out-coupling transmissivity equal to the round trip loss (see supplementary material, Sec. C). Ref. \cite{Jonathan2022} shows that a coherent states in a lossless all-pass photonic ring resonators are optimal for fixed mean input photon number. Our analysis extends those results by showing ring resonators with a homodyne readout are dose-optimal, even in the presence of loss. In the interest of minimizing optical loss and maximizing interferometric stability for microscopy, a Fabry-Perot cavity as shown in Fig. \ref{fig:schematic} (C) may be more practical. This nearly-optimal configuration corresponds to a CI with $\tau_k=\epsilon\eta^{(m-k)}$ for odd $k$ and $\tau_k=0$ otherwise (see supplementary material, Sec. D).

Fig. \ref{fig:QL} shows $\xi/\xiQL$ for each of the measurements described so far vs (A) the absorption probability $1-\eta$ and (B) the relevant measurement parameter ($m$, $n$, or $\nsq$) with $\eta=0.9$. The lines labeled CCI are $\xi/\xiQL$ for a CI with a constant, infinitesimal beamsplitter transmissivity using the optimal number of stages. The lines labeled OCI show $\xi/\xiQL$ for a CI with optimized beamsplitter transmissivities. For $\eta=0.9$, a 10(32)-stage CI (using either strategy) is more dose efficient than a standard interferometer using a Gaussian probe squeezed by 10(20) dB. The Fabry-Perot cavity (FP) with out-coupling mirror transmissivity matched to the round-trip loss is nearly optimal for all $\eta$.

\begin{figure*}
\includegraphics[width=\textwidth]{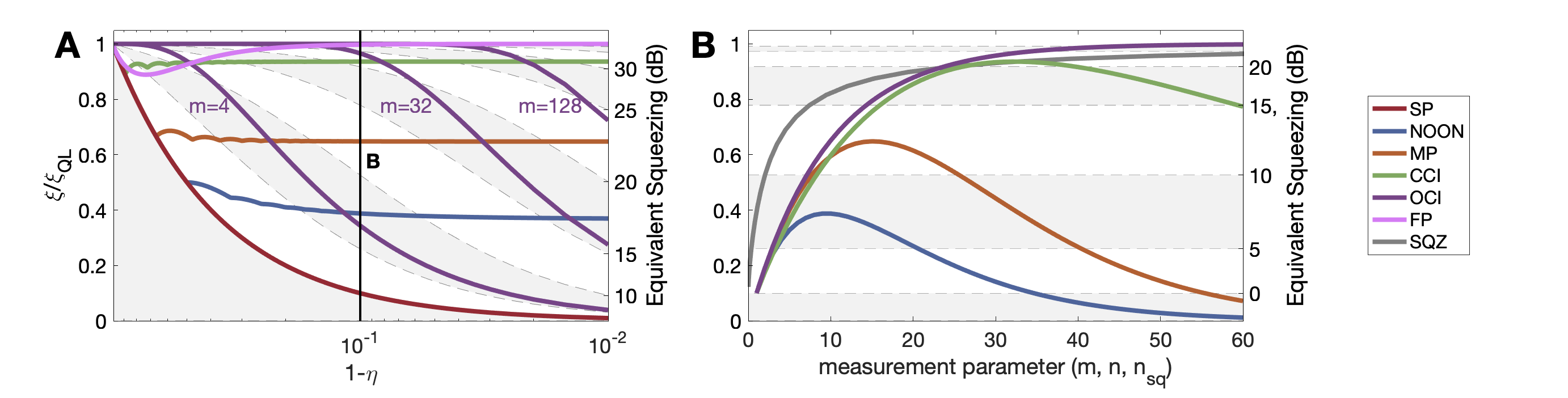}
\caption{\label{fig:QL} QFI per dose ($\xi$) relative to the quantum limit ($\xiQL$) for measurements with: a single particle or classical probe and single-pass (SP); a maximally entangled probe of size $n$ (NOON); a single particle or classical probe passed $m$ times through the sample (MP); an $m-$stage chain interferometer with constant beamsplitter angles (CCI); an $m$-stage chain interferometer with optimized beamsplitter angles (OCI); a Fabry-Perot cavity with out-coupling transmissivity equal to $\eta^2$ (FP);  and a squeezed Gaussian probe state with $\nsq$ particles contributing to the squeezing (SQZ). (A) $\xi/\xiQL$ vs sample absorbance. For the NOON, MP, and CCI, $m$ and $n$ are optimized for each $\eta$. For the OCI $m$=4, 32, or 128 stages. (B) $\xi/\xiQL$ vs the relevant measurement parameter for $\eta=0.9$ (marked by a vertical black line in (A)). The striped backgrounds in (A) and (B) show the equivalent squeezing: the squeezing (in dB) needed for a Gaussian probe to have equivalent dose efficiency.}
\end{figure*}
%%%%%%%%%%%%%%%%%%%%%%%%%%%%%%%%%%%%%%%%%%%%
\section{Lossy Optics}
%%%%%%%%%%%%%%%%%%%%%%%%%%%%%%%%%%%%%%%%%%%%

So far we have neglected losses other than those caused by the sample. We will now consider the effects of loss during probe preparation (probability $1-\etap$), loss between each stage of a MP or chain interferometer (probability $1-\etart$), and loss during detection (probability $1-\etam$). Loss during probe preparation has no effect on dose efficiency for separable probes. However it can reduce the effectiveness of entangled probes. For squeezed states, this loss injects vacuum noise, putting a ceiling on the maximum effective squeezing. For the same reason, entanglement-enhanced measurements also suffer disproportionately from loss at the detection stage. 

The dose efficiencies for each of the measurements discussed so far are derived in the supplemental document. For CI and squeezed-state measurements, the dose efficiency has the upper bound
\begin{equation}
    \xi\leq\frac{4\eta\etam}{(1-\eta\eta^*)}\ ,
\end{equation}
which is saturated in the limit $m\rightarrow \infty$ or $\nsq\rightarrow\infty$, respectively.
For a CI, $\eta^*=\etart$, and for a squeezed state interferometer, $\eta^*=\eta_P\etam$. When $\etart<\etap\etam$, the CI has superior performance. The optimal beamsplitter transmissivities $\tau_k$ are adjusted in the presence of loss as described in the supplementary material. It may be possible to build a MP interferometer with lower loss than is possible with a CI. If, for example, $\etart=0.95$ for MP, then it becomes more effective than a CI with $\etart=0.9$ for $\eta>0.96$. When $\etart$ is small, it can be advantageous to combine multi-passing and squeezing. A multi-pass squeezed-state interferometer with $\etart=0.95$ surpasses the CI (with $\etart=0.9$) for $\eta>0.8$ (see supplement).

In Fig \ref{fig:loss} we show the dose efficiency of measurements with $\etart=0.95$ and $\etap=\etam=0.9$ relative to the quantum limit when $\etart=\etap=\etam=1$. 

\begin{figure}
\includegraphics[width=.48\textwidth]{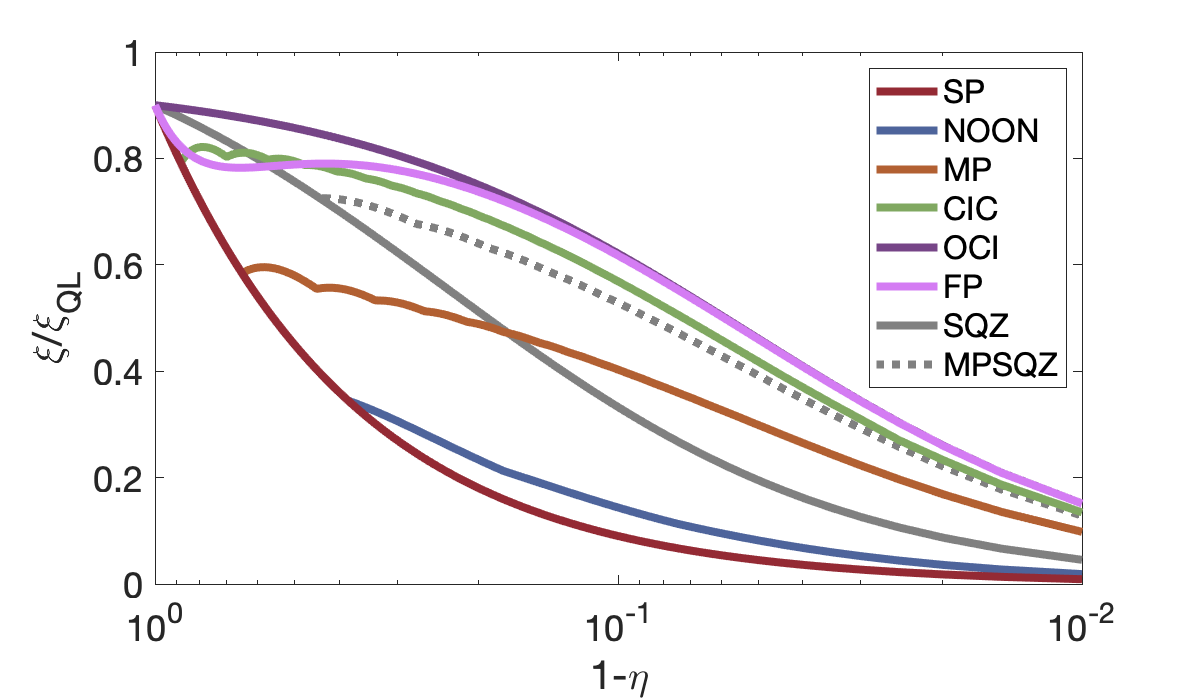}
\caption{\label{fig:loss} QFI per dose with 10\% loss during both probe preparation and detection and 5\% loss at each stage of the MP, FP, and CI measurements. The measurements are as described in Fig. \ref{fig:QL}. The curve labeled SQZ represents the upper bound on squeezing enhancement ($\nsq\rightarrow\infty$), and the dotted line labeled MPSQZ is the bound measurements with both squeezing and multiple passes.}
\end{figure}

%%%%%%%%%%%%%%%%%%%%%%%%%%%%%%%%%%%%%%%%%%%%
\section{Conclusion}
%%%%%%%%%%%%%%%%%%%%%%%%%%%%%%%%%%%%%%%%%%%%

We have identified phase estimation schemes which attain the quantum limit for information per dose without using entanglement. This makes it possible, in principle, to employ a quantum-optimal measurement of weak phase objects using only basic optical elements. The CI is compatible with short coherence time single-particle probes like the electrons. For longer coherence time probes, equivalent measurements can be performed using leaky cavities. When taking into account losses in the optics for forming the probe and performing the measurement of the exit wavefunction, we find that CI and cavity-enhanced measurements can outperform entanglement-enhanced measurements.

\bibliography{refs.bib}% Produces the bibliography via BibTeX.

\end{document}